\documentclass{sig-alternate}
\usepackage{color}
\usepackage{rotating}
\usepackage{colortbl}
\usepackage{graphics}
\usepackage{graphicx}
\usepackage{multirow}
\usepackage{latexsym}
\usepackage{amsmath}
\usepackage{lscape}
\usepackage{pgf}
\usepackage{tikz}
\usepackage{xcolor}
\usepackage{hyperref}
\usepackage{pgfplots}
\begin{document}
%
\title{Fixed versus Dynamic Co-Occurrence Windows in TextRank Term Weights for Information Retrieval}

%
%
%
%
%

\numberofauthors{3} 
%
\author{
%
%
\alignauthor
Wei Lu\\
       \affaddr{School of Information Management}\\
       \affaddr{Wuhan University, China}\\
       \email{reedwhu@gmail.com}
\alignauthor
Qikai Cheng\\
       \affaddr{School of Information Management}\\
       \affaddr{Wuhan University, China}\\
       \email{chengqikai0806@gmail.com}
\alignauthor 
Christina Lioma\\
       \affaddr{Computer Science}\\
       \affaddr{University of Copenhagen, Denmark}\\
       \email{c.lioma@diku.dk}
}
\date{30 July 1999}

\maketitle
\begin{abstract}
TextRank is a variant of PageRank typically used in graphs that represent documents, and where vertices denote terms and edges denote relations between terms. Quite often the relation between terms is simple term co-occurrence within a fixed window of $k$ terms. The output of TextRank when applied iteratively is a score for each vertex, i.e. a term weight, that can be used for information retrieval (IR) just like conventional term frequency based term weights. 

So far, when computing TextRank term weights over co-occurrence graphs, the window of term co-occurrence is always fixed. This work departs from this, and considers dynamically adjusted windows of term co-occurrence that follow the document structure on a sentence- and paragraph-level. The resulting TextRank term weights are used in a ranking function that re-ranks 1000 initially returned search results in order to improve the precision of the ranking. Experiments with two IR collections show that adjusting the vicinity of term co-occurrence when computing TextRank term weights can lead to gains in early precision.   
\end{abstract}




\section{Introduction}
Associative networks have long been used to represent units of text and their interconnecting relations \cite{MihalceaR2011}. The symbolic structures that emerge from these representations correspond to graphs, where text constituents are represented as vertices and their interconnecting relations as edges. Graph ranking algorithms, such as the TextRank \cite{MihalceaT04,MihalceaR2011} variant of PageRank, have been used successfully in keyword extraction \cite{MihalceaT04}, classification \cite{HassanMB07} and information retrieval \cite{BlancoL12} to compute term weights from graphs of individual documents, where vertices represent the document's terms, and edges represent term co-occurrence within a fixed window. Using these computations iteratively, the weight of a term can be estimated with respect to the terms that fall in its vicinity and their respective term weights. An underlying assumption in these approaches is that the vicinity of term co-occurrence is fixed for all terms. To our knowledge, there is no theoretical or intuitive basis for this assumption.

Fixed-window term co-occurrence may not be optimal for TextRank term weights. Lexical affinities may span across more words in longer sentences than they do in shorter sentences. Hence, adjusting the co-occurrence window according to the discourse span of the text might be a better choice. Based on this intuition, in this work we look at the effect of dynamically adjusted windows of term co-occurrence upon their resultant TextRank term weights. We experiment with co-occurrence windows that follow the document structure on two levels of granularity: sentences and paragraphs. For each of these, we compute term weights using TextRank, and use them for retrieval using the ranking model of \cite{BlancoL12}, i.e. linearly combined with inverse document frequency (idf). Experiments using these TextRank term weights for re-ranking the top 1000 search results show that sentence-based co-occurrence can outperform fixed-window co-occurrence in terms of early precision.

\section{Co-Occurrence Windows}
\label{s:Motivation}
\subsection{Methodology}
We experiment with two datasets: Reuters RCV1 from TREC 2002 (2.5GB, 50 title-only queries) and INEX 2005 (764MB, 47 content-only queries). We build a separate graph for each document: terms are represented as vertices (initially unweighted), and term co-occurrence within a window is represented as an undirected edge linking the vertices of the co-occurring terms. We use TextRank \cite{MihalceaT04} to compute iteratively the score of each vertex $v_i$: 
\begin{equation}
s(v_i) = (1 - \delta) + \delta \times \sum_{j \in V(v_i)} \frac{S(v_j)}{|V(v_j)|}
\end{equation} 
\noindent where $s(v_i)$ is the TextRank score of vertex $v_i$, $V(\cdot)$ denotes the set of vertices connecting with a vertex, $|\cdot |$ marks cardinality, and $0\le \delta \le 1$ is a damping factor that integrates into the computation the probability of jumping randomly from one vertex to another. We iterate the formula 200 times, using the default $\delta=0.85$ \cite{MihalceaT04}. The final score of each vertex represents a term weight where the higher the number of different words that a given word co-occurs with, and the higher their weight, the higher the weight of this word. It has been shown that a nonlinear correlation exists between such TextRank term weights and term frequency based term weights \cite{MihalceaR2011}. 

We use these term weights to compute the score of a document for a query ($s(d,q)$) according to \cite{BlancoL12}: 
\begin{equation}
s(d,q) = \sum_{i \in q} \log idf_i \times \log s(i)
\end{equation} 
\noindent where $i$ is a query term, and $s(i)$ is the corresponding TextRank score for vertex $v_i$. No document length normalisation is used. We use Porter's stemmer for the documents and queries.

To compare fixed versus dynamically adjusted windows of term co-occurrence, we use a baseline where the window of term co-occurrence is fixed to the best values reported in the IR literature (albeit for other datasets)\footnote{In non-IR literature, optimal fixed values are: $k=$2,4 for classification \cite{HassanMB07} and $k=$2 for keyword extraction \cite{MihalceaT04}, however these values consistently underperform for IR \cite{BlancoL07,BlancoL12}.}\cite{BlancoL12}: $k=$5 \& 6.  We compare this baseline against term co-occurrence that is dynamically adjusted to the length of each (a) sentence and (b) paragraph\footnote{We treat these elements as paragraphs: \texttt{p} (for RCV1) and \texttt{ilrj, ip1, ip2, ip3, ip4, ip5, item-none, p, p1, p2, p3, Bib, Bm, St} (for INEX).}, separately. The sentence/paragraph term statistics are displayed in Table \ref{tab:stats}. We evaluate this comparison in a re-ranking scenario, where the task is to re-rank an initially retrieved set of 1000 documents. For the INEX collection (where relevance assessments apply to document sections) we consider a document relevant if any of its containing sections is assessed relevant.  

\subsection{Findings}
Table \ref{tab:scores} shows different metrics of retrieval performance when using fixed versus sentence- and paragraph-length windows of term co-occurrence. We see that results vary\footnote{Results were not stat. significant when the t-test was used.}. For average precision (NDCG) fixed co-occurrence is best for RCV1, and sentence-based co-occurrence is best for INEX. The reverse happens for precision in the top 10 retrieved documents (P@10): fixed co-occurrence is best for INEX, and sentence-based co-occurrence is best for RCV1. The only consistent trend is in the precision of the single top retrieved document (MRR), which benefits more from dynamically adjusted co-occurrence consistently for both collections. This finding is novel, considering the earlier position of \cite{MihalceaT04} that the larger the window of co-occurrence, the lower the precision. This finding indicates that larger window sizes may lead to gains in precision, if however they are not fixed but rather dynamically adjusted to text units like sentences. 

Finally, sentences appear to be an overall better boundary of term co-occurrence than paragraphs, with the exception of NDCG for INEX where paragraph-based co-occurrence slightly outperforms sentence-based co-occurrence (and they both outperform fixed co-occurrence). This could be due to the fact that INEX paragraphs are relatively short and focused content-wise \cite{MalikKLF05}.

\begin{table}
\begin{center}
\resizebox{85mm}{!}{
\begin{tabular}{|l|r|r|r|r|}
\hline
&sent (RCV1)&para (RCV1)&sent (INEX)&para (INEX)\\
\hline
min length&1&1&1&1\\
max length&1731&31696&7920&111136\\
min tokens&1&1&1&1\\
max tokens&250&4662&2447&17379\\
average tokens&19.87&20.35&15.73&58.51\\
\hline
\end{tabular}
}
\caption{Sentence (sent) and paragraph (para) statistics per retrieval dataset.}
\label{tab:stats}
\end{center}
\begin{center}
\resizebox{85mm}{!}{
\begin{tabular}{|l|l||l|l|l|l|l|l|}
\hline
\multicolumn{8}{|c|}{ Re-ranking top 1000 retrieved documents}\\
\hline
\multicolumn{2}{|c||}{co-occurrence} &\multicolumn{3}{c|}{RCV1}&\multicolumn{3}{c|}{INEX}\\
\multicolumn{2}{|c||}{window}   &NDCG&MRR&P@10&NDCG&MRR&P@10\\
\hline
\multirow{2}{*}{fixed}
&5 terms		&\bf0.5238&0.6736&0.4300	&0.5541&0.6865&\bf0.4750\\
&6 terms	&0.5025&0.6559&0.4280	&0.5540&0.6966&0.4714\\
\hline
\multirow{2}{*}{dynamic}
&sentence 	&0.5119 &\bf0.6811 &\bf0.4340 	&0.5543&\bf0.7021&0.4743\\
&paragraph 	&0.5178 &0.6574 &0.4160 	&\bf0.5545&0.6975&0.4714\\
\hline
\end{tabular}
}
\caption{Retrieval performance with TextRank term weights using fixed vs. dynamic co-occurrence windows, on two datasets. Bold font marks best scores.}
\label{tab:scores}
\end{center}
\end{table}


 \section{Conclusion}
\label{s:Conclusion}
We modelled individual documents as separate graphs where vertices represent terms, and co-occurrence relations among terms represent edges. Using the TextRank model of Mihalcea et al. \cite{MihalceaT04,MihalceaR2011} we computed vertex weights corresponding to term weights, which we used for retrieval using the ranking of Blanco et al. \cite{BlancoL07,BlancoL12}. Unlike all these existing approaches where term co-occurrence is fixed to a window of $k$ terms at all times, we reasoned that term co-occurrence should be varied according to sentence or paragraph length. Our motivation was that meaningful term relations may span across more words in longer sentences than they do in shorter sentences, hence fixing term co-occurrence may not be optimal for all terms. 

Preliminary experiments in a re-ranking scenario with two retrieval datasets showed that sentence-based co-occurrence can lead to early precision gains over fixed term co-occurrence at 5 and 6 terms, which are optimal values in the IR literature. More experiments with larger datasets and full-ranking (as opposed to re-ranking) documents are needed to investigate the optimal term co-occurrence vicinity. This small-scale work contributes a novel comparison between fixed versus dynamically adjusted co-occurrence windows for TextRank term weights, and the initial finding that sentence-based co-occurrence can improve early precision.


\textbf{Acknowledgments.} We thank Birger Larsen and the anonymous reviewers for useful feedback. Work partially funded by DANIDA (grant no. 10-087721) and the National Natural Science Foundation of China (grant no. 71173164).

%
\bibliographystyle{abbrv}
\bibliography{graph}  

\begin{thebibliography}{1}

\bibitem{BlancoL07}
R.~Blanco and C.~Lioma.
\newblock Random walk term weighting for information retrieval.
\newblock In W.~Kraaij, A.~P. de~Vries, C.~L.~A. Clarke, N.~Fuhr, and N.~Kando,
  editors, {\em SIGIR}, pages 829--830. ACM, 2007.

\bibitem{BlancoL12}
R.~Blanco and C.~Lioma.
\newblock Graph-based term weighting for information retrieval.
\newblock {\em Inf. Retr.}, 15(1):54--92, 2012.

\bibitem{HassanMB07}
S.~Hassan, R.~Mihalcea, and C.~Banea.
\newblock Random walk term weighting for improved text classification.
\newblock {\em Int. J. Semantic Computing}, 1(4):421--439, 2007.

\bibitem{MalikKLF05}
S.~Malik, G.~Kazai, M.~Lalmas, and N.~Fuhr.
\newblock Overview of inex 2005.
\newblock In N.~Fuhr, M.~Lalmas, S.~Malik, and G.~Kazai, editors, {\em INEX},
  volume 3977 of {\em Lecture Notes in Computer Science}, pages 1--15.
  Springer, 2005.

\bibitem{MihalceaR2011}
R.~Mihalcea and D.~Radev.
\newblock {\em Graph-Based Natural Language Processing and Information
  Retrieval}.
\newblock Cambridge University Press, 2011.

\bibitem{MihalceaT04}
R.~Mihalcea and P.~Tarau.
\newblock Textrank: Bringing order into text.
\newblock In {\em EMNLP}, pages 404--411. ACL, 2004.

\end{thebibliography}
%
%

\end{document}